\documentclass[12pt]{article}
\pdfoutput =1

\usepackage{amsmath,amssymb}
\usepackage{slashed}
\usepackage{xcolor} 
\usepackage{graphicx}
\usepackage[utf8]{inputenc}  
\usepackage{url}
\usepackage{cite}

\renewcommand{\imath}{i} 

\usepackage[normalem]{ulem}

\usepackage{subfig}

\usepackage{hyperref}
\hypersetup{
	backref=true,            
	pagebackref=true,        
	breaklinks=true,         
	hyperindex=true,         
	hyperfootnotes=true,     
	colorlinks=true,         
	linkcolor=blue,           
	citecolor=blue, 
	filecolor=magenta,       
	urlcolor=blue,           
	runcolor=filecolor,      
	menucolor=red,           
	pdfdisplaydoctitle=true, 
	bookmarks=true,          
	bookmarksopen=false,     
	unicode=true,            
	pdftitle={Exotic Higgs}, 
	pdfwindowui=true,        
	pdftoolbar=true,         
	pdfmenubar=true,         
	pdffitwindow=false,      
	pdfstartview={Fit},      
	pdfnewwindow=true        
}

\usepackage{siunitx}
\def\lsim{\mathrel{\rlap{\lower4pt\hbox{\hskip1pt$\sim$}}
  \raise1pt\hbox{$<$}}}
\def\gsim{\mathrel{\rlap{\lower4pt\hbox{\hskip1pt$\sim$}}
  \raise1pt\hbox{$>$}}}

\newcommand{\beq}{\begin{equation}}
\newcommand{\eeq}{\end{equation}}
\newcommand{\bea}{\begin{eqnarray}}
\newcommand{\eea}{\end{eqnarray}}

\newcommand{\nnl}{\nonumber \\}
\newcommand{\gev}{{\rm GeV}}
\newcommand{\hc}{\mathrm{h.c.}}
\newcommand{\eps}{\epsilon}

\newcommand{\cO}{{\cal O}}
\newcommand{\cL}{{\cal L}}

\graphicspath{{./Figures/}}

\newcommand{\eref}[1]{Eq.~\eqref{eq:#1}}

\newcommand{\sref}[1]{Section~\ref{sec:#1}}
\newcommand{\tref}[1]{Table~\ref{tab:#1}}

\begin{document}

\vspace*{-2cm}
\begin{flushright}
LPT-Orsay-16-11 \\
\vspace*{2mm}
\today
\end{flushright}

\begin{center}
\vspace*{15mm}

\vspace{1cm}
{\large \bf
On the exotic Higgs decays in effective field theory
} \\
\vspace{1.4cm}

{Herm\`es B\'elusca-Ma\"ito\footnote{Email: hermes.belusca@th.u-psud.fr} and Adam Falkowski\footnote{Email: adam.falkowski@th.u-psud.fr}}

 \vspace*{.5cm}
Laboratoire de Physique Th\'{e}orique, Bat.~210, Universit\'{e} Paris-Sud, 91405 Orsay, France

\vspace*{.2cm}

\end{center}

\vspace*{10mm}
\begin{abstract}\noindent\normalsize
We discuss  exotic Higgs decays in an effective field theory where the Standard Model is extended by dimension-6 operators.
We review and update the status of 2-body lepton- and quark-flavor violating decays involving the Higgs boson.
We also comment on the possibility of observing 3-body flavor-violating Higgs decays in this context.

\end{abstract}
\vfill
\tableofcontents
\vfill

\vspace*{3mm}

\newpage

\section{Introduction}
\label{sec:intro}

If new particles beyond the Standard Model (SM) are much heavier than 100 GeV, physics at the weak scale can be described by an effective field theory (EFT) with the SM Lagrangian perturbed by higher-dimensional operators.
The latter encode, in a model-independent way, possible effects of new heavy  particles at energies well below the new physics scale $\Lambda$.
The EFT framework allows for a systematic expansion of these effects in operator dimensions or, equivalently, in powers of $1/\Lambda$.
The leading effects are expected from operators of dimension 6, as their coefficients are suppressed by $1/\Lambda^2$.
The first classification of dimension-6 operators was performed in Ref.~\cite{Buchmuller:1985jz}.
For 1 generation of fermions, a complete non-redundant set (henceforth referred to as a {\em basis}) was identified and explicitly written down in Ref.~\cite{Grzadkowski:2010es}.
Ref.~\cite{Alonso:2013hga} extended this to 3 generations of fermions, in which case a basis is characterized by 2499 independent parameters.


In this paper we are interested in the subset of these operators that lead to exotic decays of the 125 GeV Higgs boson.
By ``exotic'' we mean decays that are forbidden in the SM or predicted to occur with an extremely suppressed branching fraction.
More specifically, we are interested in decays that violate the lepton flavor or quark flavor.
Lepton flavor violating (LFV) processes are completely forbidden in the SM in the limit of zero neutrino masses.
Quark flavor violating (QFV) Higgs decays as a flavor-changing neutral current process  are forbidden in the SM at tree-level.
Theoretical studies of exotic Higgs decays have a long history, see e.g. Refs.~\cite{Strassler:2006ri,Strassler:2006im,Gopalakrishna:2008dv,Davoudiasl:2012ag,Davoudiasl:2013aya,Huang:2013ima,Gonzalez-Alonso:2014rla,Falkowski:2014ffa} and  \cite{Curtin:2013fra} for a review.
Most of these papers assume new light degrees of freedom, in which case the EFT approach described here is not adequate.
On the other hand, Refs.~\cite{Blankenburg:2012ex,Harnik:2012pb} recently studied the possibility of LFV and QFV 2-body decays of the 125~GeV Higgs within the EFT framework.
Such decays can arise in the presence of Yukawa-type dimension-6 operators  \cite{DiazCruz:1999xe}.
These papers demonstrated that  LFV Higgs decays to $\tau^\pm \mu^\mp$ and $\tau^\pm e^\mp$ with the branching fraction as large as 10\% are allowed by current indirect constraints. At the same time, the LHC is currently sensitive to branching fractions of order 1\% \cite{Chatrchyan:2014tja}.
This corresponds to probing the scale suppressing the corresponding dimension-6 operators at the level of $\Lambda \sim 10$~TeV.

The goal for this paper is to extend this study to a full set of dimension-6 operators.
Apart from the Yukawa-type operators, exotic Higgs decays can arise in the presence of vertex-type $\sim H^\dagger H \bar \psi \gamma_\mu \psi$ and dipole-type $\sim H \bar \psi \sigma^{\mu \nu} \psi F_{\mu \nu}$ operators.
We systematically discuss these operators and the new Higgs decay channels that they imply.
The structure of the dimension-6 Lagrangian then implies certain relations between these Higgs couplings, as well as relations between single-Higgs interactions and  Lagrangian terms without a Higgs  that affect precision observables.
We give the limits  on each of the couplings from precision tests of the SM.
That information can be explored to place limits on the allowed magnitude of the Higgs couplings.
We will discuss the maximum exotic Higgs branching fraction that these limits permit.

The paper is organized as follows.
In \sref{d6} we define our notation and introduce the dimension-6 Lagrangian with LFV and QFV interactions in the Higgs basis.
In \sref{2b} we review and update the results of Refs.~\cite{Blankenburg:2012ex,Harnik:2012pb} concerning two body exotic Higgs decays.
In \sref{3b} we study the possibility of LFV and QFV Higgs decays mediated by vertex- and dipole-type operators, respectively.
Obviously, studying the full parameter space of the dimension-6 Lagrangian would be an extremely difficult task.
To deal with the degeneracies among the parameters, one simplifying assumption we make throughout this paper is that the flavor-diagonal Higgs couplings are not significantly affected by higher-dimensional operators.
Furthermore, we will assume that there are no large fine-tuned cancellations between different parameters so as to satisfy constraints from precision experiments.
In such a constrained framework, we discuss the limits on the LFV and QFV Higgs couplings from various precision measurements.
Given these constraints, we discuss the implications for the rate of exotic Higgs decays at the LHC.
%

\section{Exotic Higgs couplings from dimension-6 Lagrangian}
\label{sec:d6}

We consider an effective theory where the SM is extended by dimension-6 operators:
\beq
\label{eq:D6_def}
\cL_{\rm eff} = \cL^{\rm SM} + \frac{1}{v^2} \cL^{D=6}.
\eeq
We assume the SM electroweak symmetry is {\em linearly} realized.
This implies $\cL_{\rm eff}$ contains local operators invariant under the $SU(3) \times SU(2) \times U(1)$ symmetry;
in particular, the Higgs boson $h$ enters the Lagrangian only through gauge invariant interactions of the Higgs doublet $H$.
The SM Lagrangian in our notation takes the form:
\begin{equation}\begin{split}
\label{eq:lsm}
{\cal L}^{\rm SM} =\;&
  -\frac{1}{4} G^a_{\mu\nu} G^{a\;{\mu\nu}} - \frac{1}{4} W^i_{\mu\nu} W^{i\;{\mu\nu}} - \frac{1}{4} B_{\mu\nu} B^{\mu\nu}
  + D^\mu H^\dagger D_\mu H + \mu_H^2 H^\dagger H - \lambda (H^\dagger H)^2 \\
& + \sum_{f \in q, \ell} i \bar f_L \gamma^\mu D_\mu f_L + \sum_{f \in u ,d,e} i \bar f_R \gamma^\mu D_\mu f_R \\
& - \left[ \tilde H^\dagger \bar u_R Y^u q_L + H^\dagger \bar d_R Y^d V_{\rm CKM} q_L + H^\dagger \bar e_R Y^\ell \ell_L + \hc \right].
\end{split}\end{equation}
The gauge couplings of $SU(3) \times SU(2) \times U(1)$ are denoted by $g_S$, $g_L$, $g_Y$, respectively; we also define the electromagnetic coupling $e = g_L g_Y/\sqrt{g_L^2 + g_Y^2}$, and the weak angle $s_\theta = g_Y/\sqrt{g_L^2 + g_Y^2}$.
The Higgs doublet $H$ acquires the VEV $\langle H \rangle = (0,v/\sqrt{2})$, where $v \approx 246.2$~GeV.
We also define $\tilde H_i = \eps_{ij} H_j^*$.
After electroweak symmetry breaking, the gauge mass eigenstates are defined as
$W^\pm = (W^1 \mp i W^2)/\sqrt 2$, $Z = c_\theta W^3 - s_\theta B$, $A = s_\theta W^3 + c_\theta B$, where $c_\theta = \sqrt{1 - s_\theta^2} = g_L/\sqrt{g_L^2 + g_Y^2}$.
The fermions $q_L = ( u_L , V_{\rm CKM}^\dagger  d_L)$ and $\ell_L = (\nu_L, e_L)$ are doublets of the $SU(2)$ gauge group.
All fermions are 3-component vectors in the generation space.
We work in the basis where the fermions are mass eigenstates, thus $Y^{u,d,\ell}$ are $3 \times 3$ diagonal matrices such that: $[Y^f]_{ij} \frac{v}{\sqrt{2}} = m_{f_i} \delta_{ij}$.
The Higgs boson interactions following from \eref{lsm},
\beq
\label{eq:SMh}
\cL^{\rm SM}_h =
 \left( \frac{h}{v} + \frac{h^2}{2 v^2} \right)  \left[ 2 m_W^2  W^+_\mu W^{-\;\mu} + m_Z^2 Z_\mu Z^\mu \right]
 -  \frac{h}{v} \sum_f m_f \bar f f - \frac{m_h^2}{2 v} h^3 - \frac{m_h^2}{8 v^2} h^4 ,
\eeq
do not contain any LFV nor QFV couplings.


We move to describe the effect of dimension-6 operators.
In \eref{D6_def} we choose to normalize them by the electroweak scale $v$, while the new physics scale $\Lambda$ is absorbed into the coefficients $c_i \sim v^2/\Lambda^2$ of these operators in the Lagrangian.
A complete non-redundant $\cL^{D=6}$ for 3 generations of fermions was explicitly written down in Ref.~\cite{Alonso:2013hga}.
Here we work at the level of Higgs boson couplings with other SM mass eigenstates, as in \cite{Gupta:2014rxa,HXSWGbasis}.
In this language, the Lagrangian is defined by a set of couplings $[\delta y_f]$, $[\delta g^{Vf}]$, and $[d_{Vf}]$, which are in general $3\times 3$ matrices with non-diagonal elements for all fermion species $f$.
A subset of these interactions violates lepton flavor and introduces tree-level flavor changing neutral currents for quark flavor violation.

The first group is related to corrections to the SM Higgs Yukawa couplings in \eref{SMh}:
\beq
\label{eq:hff}
\cL_{hff}^{D=6} = - \frac{h}{v} \sum_{f \in u,d,e} \sum_{i \neq j}
\sqrt{m_{f_i} m_{f_j}} \left[ [\delta y_f]_{ij} \bar f_{R,i} f_{L,j} + \hc \right] .
\eeq
These couplings arise from dimension-6 operators of the form $c_f |H|^2 \bar{f} H f$, with $[c_f]_{ij} \sim \sqrt{m_{f_i} m_{f_j}} [\delta y_f]_{ij}$.

The second group is related to the contact interactions between the Higgs boson, fermions, and the massive $SU(2)$ vector bosons:
\begin{equation}\begin{split}
\label{eq:hvff}
	\cL_{hVff}^{D=6} =\;&
		\frac{g_L}{\sqrt{2}} \left(1 + \frac{h}{v} \right)^2  W_\mu^+ \sum_{i \neq j} \left(
		     \bar{u}_{L,i} \gamma^\mu [\delta g^{Wq}_L]_{ij} d_{L,j}
		 +   \bar{u}_{R,i} \gamma^\mu [\delta g^{Wq}_R]_{ij} d_{R,j}
		 + \bar{\nu}_{L,i} \gamma^\mu [\delta g^{W\ell}_L]_{ij} e_{L,j} \right) + \hc \\
		&+ \sqrt{g_L^2 + g_Y^2} \left(1 + \frac{h}{v} \right)^2 Z_\mu
		    \sum_{ij} \left[ \sum_{f \in u,d,e,\nu} \bar{f}_{L,i} \gamma^\mu [\delta g^{Zf}_L]_{ij} f_{L,j} +
		    \sum_{f \in u,d,e} \bar{f}_{R,i} \gamma^\mu [\delta g^{Zf}_R]_{ij} f_{R,j} \right]
\end{split}\end{equation}
where $[\delta g^{Vf}]$ are Hermitian matrices.
These couplings arise from dimension-6 operators of the form $H^\dagger D_\mu H \bar f \gamma^\mu f$.
The gauge symmetry of the dimension-6 Lagrangian implies
$\delta g^{Wq}_L = \delta g^{Zu}_L V_{\rm CKM} - V_{\rm CKM} \delta g^{Zd}_L$ and
$\delta g^{W\ell}_L = \delta g^{Z\nu}_L - \delta g^{Ze}_L$.
Furthermore, it implies that the Higgs boson enters via  $(1 + h/v)^2$.
Therefore, the strength  of the Higgs contact interactions of this form is correlated with vertex corrections to the $W$ and $Z$ boson interactions with fermions.

Finally, we also consider the dipole-type Higgs interactions:
\begin{equation}\begin{split}
\label{eq:hffdv}
	\cL_{\rm dipole}^{D=6} =\;&
		- \frac{1 + h/v}{v^2} \sum_{i \neq j} \left[
		    g_S \sum_{f \in u,d} \sqrt{m_{f_i} m_{f_j}} \bar{f}_{R,i} \sigma^{\mu\nu} T^a [d_{Gf}]_{ij} f_{L,j} G_{\mu\nu}^a
		\right. \\
		&\left.
		+ e \sum_{f \in u,d,e} \sqrt{m_{f_i} m_{f_j}} \bar{f}_{R,i} \sigma^{\mu\nu} [d_{A f}]_{ij} f_{L,j} A_{\mu\nu}
		+ \sqrt{g_L^2 + g_Y^2} \sum_{f \in u,d,e} \sqrt{m_{f_i} m_{f_j}} \bar{f}_{R,i} \sigma^{\mu\nu} [d_{Zf}]_{ij} f_{L,j} Z_{\mu\nu}
		\right. \\
		&\left.
		+ \sqrt{2} g_L \left(
		  \sqrt{m_{u_i} m_{u_j}} \bar{u}_{R,i} \sigma^{\mu\nu} [d_{Wu}]_{ij} d_{L,j} W_{\mu\nu}^+
		+ \sqrt{m_{d_i} m_{d_j}} \bar{d}_{R,i} \sigma^{\mu\nu} [d_{Wd}]_{ij} u_{L,j} W_{\mu\nu}^-
		\right)
		\right. \\
		&\left.
		+ \sqrt{2} g_L \left( \sqrt{m_{e_i} m_{e_j}} \bar{\nu}_{L,i} \sigma^{\mu\nu} [d_{We}]_{ij} e_{R,j} W_{\mu\nu}^+ \right) \right] + \hc \, ,
\end{split}\end{equation}
where $\sigma_{\mu\nu} = \frac{\imath}{2}[\gamma_\mu, \gamma_\nu]$, and $[d_{Vf}]$ are general $3 \times 3$ matrices.
These couplings are absent in the SM at the tree level, but they arise from dimension-6 operators of the form $H \bar f \sigma^{\mu \nu} f V_{\mu \nu}$.
The gauge symmetry of the dimension-6 Lagrangian implies  that the $W$ boson dipole couplings are related to those of the $Z$ boson and the photon:
$\eta_f d_{Wf} = d_{Zf} + s_\theta^2 d_{Af}$,
$\eta_u = 1$, $\eta_{d,e} = -1$.
Again, it also dictates that the Higgs boson enters via $(1 + h/v)$.
Therefore, the strength of this type of Higgs interactions is correlated with the strength of dipole interactions of the SM fermions and gauge bosons.

In \eref{hff} and \eref{hffdv} we isolated the factor $\sqrt{m_{f_i} m_{f_j}}$ in the Yukawa and dipole interactions.
This is done for convenience, and we do not assume any particular pattern of $[\delta y_f]_{ij}$ and $[d_{Vf}]_{ij}$.
The Yukawa and dipole interactions are distinguished by the fact that they violate chirality (they allow for transitions of left-handed fermions into right-handed ones and vice-versa), much like the fermion mass terms in the SM.
Any model addressing the flavor problem and generating these parameters in the low-energy EFT is expected to exhibit some sort of chiral suppression.
Exactly this pattern will arise from models following the minimal flavor violation paradigm,
where all sources of flavor violation are proportional to the SM Yukawa matrices.
Although, more generally, the chiral suppression does not have to be proportional to the fermion masses,
isolating the mass factor leads to a more transparent picture for natural values of these parameters.
For the Yukawa interactions, the off-diagonal couplings can be more readily compared to the diagonal ones which, in this normalization, are just equal to 1 in the SM limit.

In the rest of this paper, we discuss LFV and QFV exotic Higgs decays induced by the operators in \eref{hff}, \eref{hvff}, and \eref{hffdv}.
As mentioned before, we assume that the flavor-diagonal Higgs couplings are not significantly affected by higher-dimensional operators\footnote{See e.g. \cite{deLima:2015pqa} for a discussion of $D$=8 operators in this context.}, and that there are no large fine-tuned cancellations between different parameters so as to satisfy constraints from precision experiments.
In such a constrained framework, we discuss the limits on the LFV and QFV Higgs couplings from various precision measurements.
With these assumptions, we give the limits on the couplings from precision experiments and discuss the maximum exotic Higgs branching fractions allowed.

\section{Two body Higgs decays}
\label{sec:2b}

In this section we discuss two-body flavor-violating decays involving the Higgs boson.
Such processes are generated via the Yukawa couplings in \eref{hff}.
The important point is that the $[\delta y_f]_{ij}$ are free parameters from the EFT point of view, and can take any value within the EFT validity range.

\subsection{Lepton-flavor violating decays}

No experimental dedicated searches have been done so far for $h \to \mu e$ and $h \to \tau e$.
For $h \to \tau \mu$, the 95\% CL upper limit on the branching ratio was set by CMS \cite{Khachatryan:2015kon} and ATLAS \cite{Aad:2015gha}:
\beq
{\rm Br}(h \to \tau \mu) \leq 1.51\% \quad ({\rm CMS})
\quad ; \quad
{\rm Br}(h \to \tau \mu) \leq 1.85\% \quad ({\rm ATLAS}) \, .
\eeq
The CMS search shows a $2.4\sigma$ excess over the expected null background,\footnote{%
This excess may possibly be related to another one observed in the same sign di-muon final state in the $t\bar t h$  searches in ATLAS and CMS \cite{Bhattacherjee:2015sia}.}
${\rm Br}(h \to \tau \mu) = \left(0.84^{+0.39}_{-0.37} \right) \%$,
while the ``excess'' in ATLAS is only $1\sigma$,
${\rm Br}(h \to \tau \mu) = \left(0.77^{+0.62}_{-0.62} \right) \%$.
A naive combination of the ATLAS and CMS results yields:
\beq
{\rm Br}(h \to \tau \mu) = \left(0.82^{+0.33}_{-0.32} \right)\%
\quad ; \quad
{\rm Br}(h \to \tau \mu) \leq 1.47\% \quad ({\rm ATLAS+CMS}) \, .
\eeq

In terms of the parameters in \eref{hff}, the branching ratio can be written as:
\beq
\frac{{\rm Br}(h \to \tau \mu)}{{\rm Br}(h \to \tau \tau)} = \frac{m_\mu}{m_\tau} \left(
| [\delta y_\ell]_{\mu\tau} |^2  + | [\delta y_\ell]_{\tau\mu} |^2 \right),
\eeq
where we assumed the $h \to \tau \tau$ decay is not significantly affected by new physics.
Using $m_\mu = \SI{105.7}{MeV}$, $m_\tau = \SI{1.78}{GeV}$,
${\rm Br}(h \to \tau \tau) = 6.3\%$ from the SM value, we obtain the best fit value and the 95\% CL bound on the EFT parameters:
\bea
| [\delta y_\ell]_{\mu\tau} |^2 + | [\delta y_\ell]_{\tau\mu} |^2 &=& 2.19^{+0.88}_{-0.85},
\nnl
\sqrt{| [\delta y_\ell]_{\mu\tau} |^2 + | [\delta y_\ell]_{\tau\mu} |^2} &\leq& 1.98.
\eea


The strongest constraints on the LFV Higgs couplings come from $\ell_2 \to \ell_1 \gamma$ decays \cite{Blankenburg:2012ex,Harnik:2012pb}.
In the SM, such processes are completely forbidden in the limit of zero neutrino masses,
but they can be generated in the presence of $D$=6 operators.
In the EFT with LFV Yukawa couplings, they occur at one-loop level.
The amplitude for the process is parametrized as:
\begin{equation}
	\mathcal{M} = \overline{u(\ell_1)} F_2 \sigma^{\mu\nu} k_\nu u(\ell_2) \epsilon^\star_\mu(k) \quad \text{with:} \quad F_2 = \frac{1}{16\pi^2} (C_L P_L + C_R P_R) \quad ; \quad P_{L/R} = \frac{1 \mp \gamma_5}{2} \, ,
\end{equation}
and the decay width is given by (in the approximation $m_{\ell_1} \ll m_{\ell_2}$):
\begin{equation}
	\label{eq:width}
	\Gamma_{\ell_2 \to \ell_1 \gamma} \approx \frac{m_{\ell_2}^3}{4096 \pi^5}
	    \left( |C_L|^2 + |C_R|^2 \right) .
\end{equation}
Evaluating the 1-loop diagrams we find the following results:
\begin{itemize}
\item \uline{$\mu \to e \gamma$:}
\beq
\begin{pmatrix} C_L \\ C_R \end{pmatrix} \approx
    e \frac{m_\tau^2 \sqrt{m_\mu m_e}}{2 m_H^2 v^2}
    \left(2 \ln\left(\frac{m_H^2}{m_\tau^2}\right) - 3\right) \begin{pmatrix} [\delta y_\ell]_{e \tau} [\delta y_\ell]_{\tau \mu} \\ [\delta y_\ell]_{\mu \tau}^* [\delta y_\ell]_{\tau e}^* \end{pmatrix} \approx
    5.4 \times 10^{-11} \gev^{-1} \begin{pmatrix} [\delta y_\ell]_{e \tau} [\delta y_\ell]_{\tau \mu} \\ [\delta y_\ell]_{\mu \tau}^* [\delta y_\ell]_{\tau e}^* \end{pmatrix} ,
\eeq
\item \uline{$\tau \to e \gamma$:}
\beq
\begin{pmatrix} C_L \\ C_R \end{pmatrix} \approx
    e \frac{m_\tau^2 \sqrt{m_e m_\tau}}{3 m_H^2 v^2}
    \left(3 \ln\left(\frac{m_H^2}{m_\tau^2}\right) - 4\right) \begin{pmatrix} [\delta y_\ell]_{e \tau} \\ [\delta y_\ell]_{\tau e}^* \end{pmatrix} \approx
    2.2 \times 10^{-10} \gev^{-1} \begin{pmatrix} [\delta y_\ell]_{e \tau} \\ [\delta y_\ell]_{\tau e}^* \end{pmatrix} ,
\eeq
\item \uline{$\tau \to \mu \gamma$:}
\beq
\begin{pmatrix} C_L \\ C_R \end{pmatrix} \approx
    e \frac{m_\tau^2 \sqrt{m_\mu m_\tau}}{3 m_H^2 v^2}
    \left(3 \ln\left(\frac{m_H^2}{m_\tau^2}\right) - 4\right) \begin{pmatrix} [\delta y_\ell]_{\mu \tau} \\ [\delta y_\ell]_{\tau \mu}^* \end{pmatrix} \approx
    3.2 \times 10^{-9} \gev^{-1} \begin{pmatrix} [\delta y_\ell]_{\mu \tau} \\ [\delta y_\ell]_{\tau \mu}^* \end{pmatrix} .
\eeq
\end{itemize}
Above, we kept only the contributions from diagrams with the $\tau$ lepton in the internal fermion line.
Other contributions are suppressed by $m_\mu/m_\tau$ or $m_e/m_\tau$ and can be neglected, unless there is a huge hierarchy between different off-diagonal elements of $[\delta y_f]$.
Such hierarchy is not expected for EFT arising as low-energy approximation of specific models where the flavor problem is addressed.
Our results agree with Refs.~\cite{Blankenburg:2012ex,Harnik:2012pb}.

It was pointed out in the literature~\cite{Blankenburg:2012ex,Harnik:2012pb,Crivellin:2014cta,Davidson:2016utf} that certain two-loop corrections, the so-called Barr-Zee diagrams with a $W$ or a top loop, may give comparable contributions as the one-loop diagrams computed above.
Their analytical form can be found in the Appendix~A.2 of~\cite{Harnik:2012pb}, which were adapted from the $\mu \to e \gamma$ formulas of Chang~et~al.~\cite{Chang:1993kw} and Leigh~et~al.~\cite{Leigh:1990kf}.
It turns out that Barr-Zee contributions are proportional to $\sqrt{m_i m_j} \delta y_{ij} C$, where $C$ is common for all the processes.
Numerically, one finds:
\begin{itemize}
\item \uline{$\mu \to e \gamma$:}
\bea
\begin{pmatrix} C_L \\ C_R \end{pmatrix} \approx 2.3 \times 10^{-10} \gev^{-1}
    \begin{pmatrix} [\delta y_\ell]_{e \mu} \\ [\delta y_\ell]_{\mu e}^* \end{pmatrix} .
\eea
\item \uline{$\tau \to e \gamma$:}
\bea
\begin{pmatrix} C_L \\ C_R \end{pmatrix} \approx 9.6 \times 10^{-10} \gev^{-1}
    \begin{pmatrix} [\delta y_\ell]_{e \tau} \\ [\delta y_\ell]_{\tau e}^* \end{pmatrix} .
\eea
\item \uline{$\tau \to \mu \gamma$:}
\bea
\begin{pmatrix} C_L \\ C_R \end{pmatrix} \approx 1.4 \times 10^{-8} \gev^{-1}
    \begin{pmatrix} [\delta y_\ell]_{\mu \tau} \\ [\delta y_\ell]_{\tau \mu}^* \end{pmatrix} .
\eea
\end{itemize}
Indeed, the 2-loop contributions turn out to be dominant, for $\tau \to \mu \gamma$ and $\tau \to e \gamma$ by approximately a factor of $4$.
For $\mu \to e \gamma$ the ratio of two- and one-loop contributions depends on the ratios of the different off-diagonal Yukawa couplings.

	\begin{table}[tb]
		\centering
		\caption{Experimental $90\%$  CL. upper limits on the branching fraction $Br$ for lepton radiative flavour-violating processes.}
		\label{tab:brlims_rad_l_fv_processes}

		\renewcommand*{\arraystretch}{1.3}

		\begin{tabular}{ c | c | c }
			\hline\hline
			Process               & Upper limits on $Br$ & Ref./Exp. \\
			\hline
			$\mu  \to   e \gamma$ & \num{5.7e-13} & \cite{Adam:2013mnn} (MEG) \\
			$\tau \to   e \gamma$ & \num{3.3e-8}  & \cite{Aubert:2009ag} (BaBar) \\
			$\tau \to \mu \gamma$ & \num{4.4e-8}  & \cite{Aubert:2009ag} (BaBar) \\
			\hline\hline
		\end{tabular}
	\end{table}

The experimental limits on these processes obtained by the BaBar collaboration ($\tau \to \ell \gamma$), and the MEG experiment ($\mu \to e \gamma$) are collected in \tref{brlims_rad_l_fv_processes}.
Using those, we find the following constraints on the lepton-flavor violating Yukawa couplings:
\begin{itemize}
\item  \uline{$\mu \to e \gamma$:}
\beq
\sqrt{  \left| [\delta y_\ell]_{ e \mu}  + 0.2   [\delta y_\ell]_{e \tau}  [\delta y_\ell]_{\tau \mu}   \right|^2 +  \left| [\delta y_\ell]_{ \mu e} +   0.2   [\delta y_\ell]_{\mu \tau}  [\delta y_\ell]_{\tau e}  \right|^2 }
\leq 0.048.
\eeq
\item  \uline{$\tau \to e \gamma$:}
\beq
\sqrt{  \left| [\delta y_\ell]_{ e \tau} \right|^2 +  \left| [\delta y_\ell]_{ \tau e} \right|^2 }
\leq  109 .
\eeq
\item  \uline{$\tau \to \mu \gamma$:}
\beq
\sqrt{  \left| [\delta y_\ell]_{ \mu \tau} \right|^2 +  \left| [\delta y_\ell]_{ \tau \mu} \right|^2 }
\leq  8.7 .
\eeq
\end{itemize}
Limits on the off-diagonal Yukawa couplings from their one-loop contribution to $\ell_2 \to 3 \ell_1$ decays are weaker~\cite{Pruna:2015jhf}.

Finally, motivated by the constraints discussed above, we write the LFV Higgs branching fractions as:
\bea
{\rm Br}(h \to \tau \mu)  &\approx&
 { | [\delta y_\ell]_{\mu\tau} |^2  + | [\delta y_\ell]_{\tau\mu } |^2  \over 2^2  } \times 1.5 \% ,
\nnl
{\rm Br}(h \to \tau e)  & \approx &
{ | [\delta y_\ell]_{e \tau} |^2  + | [\delta y_\ell]_{\tau e} |^2  \over 100^2  }  \times 18 \% ,
\nnl
{\rm Br}(h \to \mu e)  & \approx &
{ | [\delta y_\ell]_{e \mu} |^2  + | [\delta y_\ell]_{\mu e} |^2  \over 0.06^2   } \times 4 \times 10^{-9}.
\eea
We can immediately see that the indirect constraints allow for a sizable branching fraction of
$h \to \tau e$, and $h \to \tau \mu$ decays.
In particular, the percent-level branching fraction for $h \to \tau \mu$, hinted at by the CMS excess,
can be addressed in the EFT context without any tension with $\tau \to \mu \gamma$ bounds.
However, one should note that the $\mu \to e \gamma$ constraint does not allow
${\rm Br} (h \to \tau \mu)$ and ${\rm Br} (h \to \tau e)$ to be {\em simultaneously} large.
Observing both of these decays at the LHC would thus signify a breakdown of the EFT approach.
On the other hand, ${\rm Br}(h \to \mu e)$ is constrained to be small by the $\mu \to e \gamma$ constraint, so as to be unobservable in practice.

There is also the question which explicit BSM models may generate the pattern of LFV Yukawa couplings required to produce ${\rm Br} (h \to \tau \mu/e)$ at the level of a percent to per-mille.
This turns out to be difficult in concrete models.
Typically, satisfying all constraints is either completely impossible~\cite{Falkowski:2013jya}, or requires
some fine-tuning and/or challenging model building~\cite{Dery:2014kxa,Sierra:2014nqa,deLima:2015pqa,Crivellin:2015mga,Crivellin:2015lwa,Crivellin:2015hha,Varzielas:2015joa,Aloni:2015wvn,He:2015rqa, Baek:2015mea,Dorsner:2015mja,Bizot:2015qqo,Buschmann:2016uzg}.

\subsection{Flavor changing top quark decays}

Dimension-6 operators may also violate flavor in the quark sector.
In the SM, quark flavor is not conserved due to off-diagonal CKM matrix elements,
but flavor-changing neutral currents are forbidden at tree level.
Therefore, the quark flavor violating processes involving the Higgs boson are suppressed by a loop factor, and in addition suppressed by the GIM mechanism.
On the other hand, the couplings in \eref{hff} may lead to flavor-changing neutral currents at tree level.

From the experimental point of view, the most interesting of these processes are the ones involving the top quark.
ATLAS and CMS have performed direct searches for Higgs-mediated flavor-changing neutral currents in top quark decays: $t \to h q$, $q = c, u$.
Due to loop and GIM suppression, the branching fractions for these decays in the SM are prohibitively small.
However in models beyond the SM with new sources of flavor violation these decays are often enhanced to a level that may be observable at the LHC, see e.g.~\cite{Greljo:2014dka}.

In the limit of massless charm or up quarks, the tree-level decay width is given by the formula:
\beq
\Gamma(t \to h q ) = {m_t^2 m_q \over 32 \pi v^2} \left(1 - {m_h^2 \over m_t^2} \right)^2
 \left( | [\delta y_u]_{q t} |^2  + | [\delta y_u]_{t q} |^2 \right).
\eeq
This translates to the branching fractions:
\bea
Br(t \to h c)  & = & 1.1 \times 10^{-3} \left( | [\delta y_u]_{c t} |^2  + | [\delta y_u]_{t c} |^2 \right),
\nnl
Br(t \to h u)  & = & 1.9  \times 10^{-6} \left( | [\delta y_u]_{u t} |^2  + | [\delta y_u]_{t u} |^2 \right),
\eea
where we used $\Gamma_t \approx 1.35$~GeV.

The current $95\%$ upper limits on the branching fractions for these decays are given
in \tref{brlims_Higgs_qfv_processes}.
Using these, we find the following constraints on the off-diagonal Higgs Yukawa couplings:
\bea
\label{eq:yhtc_constraints}
\sqrt{ | [\delta y_u]_{c t} |^2  + | [\delta y_u]_{t c} |^2}  & \leq & 2.1,
\nnl
\sqrt{ | [\delta y_u]_{u t} |^2  + | [\delta y_u]_{t u} |^2}  & \leq & 49.
\eea

\begin{table}[tb]
	\centering
	\caption{List of experimental $95\%$~CL. upper limits on the branching fraction $Br$ for Higgs-mediated quark flavour-violating processes.}
	\label{tab:brlims_Higgs_qfv_processes}

	\renewcommand*{\arraystretch}{1.3}

	\begin{tabular}{ c | c | c }
		\hline\hline
		Process           & Upper limits on $Br$  & Ref. \\
		\hline
		$t \to c h$ & \num{4.6e-3} (ATLAS)  & \cite{Aad:2015pja} \\
		$t \to u h$ & \num{4.5e-3} (ATLAS)  & \cite{Aad:2015pja} \\
		$t \to q(=c+u) h$ & \num{7.9e-3} (ATLAS)  & \cite{Aad:2014dya} \\
		$t \to c h$       & \num{5.6e-3} (CMS)    & \cite{CMS:2014qxa} \\ 
		\hline\hline
	\end{tabular}
\end{table}

Much as for LFV Higgs decays to tau leptons, the current indirect constraints on $\delta y_{qt}$ and $\delta y_{tq}$ do not forbid the $t \to h q$ branching fraction to be close to the current LHC limits.
While the relative phase between $\delta y_{qt}$ and $\delta y_{tq}$ is severely constrained by neutron electric dipole moment searches~\cite{Gorbahn:2014sha}, the absolute values (which enter into the $t \to h q$ widths) are allowed to be large.
One should also mention that $D$-meson oscillations place more severe constraints on the products
$\delta y_{ut} \delta y_{tc}$ and $\delta y_{tu} \delta y_{ct}$, see~\cite{Gorbahn:2014sha}.
Therefore, in the EFT context, it is impossible for {\rm both}
$t \to h c$ and $t \to h u$ branching fractions to be close to the current experimental limits.

\section{3-body decays}
\label{sec:3b}

In the previous section we discussed two body exotic decays induced by dimension-6 operators of the Yukawa type.
We concluded that indirect constraints on the LFV and QFV Higgs Yukawa couplings to fermions are consistent with the branching fractions of $h \to \tau \mu$ and $h \to \tau e$ decays that are readily observable at the LHC.
In fact, the best limits on the relevant couplings currently come from the LHC.
This agrees with conclusions from previous literature~\cite{Harnik:2012pb}.
In this section we extend this discussion to 3-body exotic Higgs decays and other operators appearing at the dimension-6 level in the EFT Lagrangian.

\subsection{$h \to W b q$}

We begin with the $h \to t^* q \to W b q$ decays.
These decays are mediated by the same Yukawa couplings that lead to the $t \to h c/u$ decays,
and are constrained by ATLAS and CMS searches as in \eref{yhtc_constraints}:
\bea
{\rm Br} (h \to W b c) & = & 1.3 \times 10^{-4} \left( | [\delta y_u]_{c t} |^2  + | [\delta y_u]_{t c} |^2 \right),
\nnl
{\rm Br} (h \to W b u) & = & 2.3 \times 10^{-7} \left( | [\delta y_u]_{u t} |^2  + | [\delta y_u]_{t u} |^2 \right),
\eea
where we summed over the $W^+$ and $W^-$ modes.
Note that Higgs decays with $\cO(10^{-4})$ branching fractions have {\em already} been seen in LHC Run-1 in the $h \to ZZ \to 4 \ell$ channel.
Thus, if $t \to h q$ decays are observed at the LHC close to the current limit, it should be possible to also observe the $h \to W b q$ decays in the future (although the $t \bar t$ background will be a challenge in this case).

\subsection{$h \to \ell_1 \ell_2 \gamma$}

We move to the dipole-type operators in \eref{hffdv}.
In the lepton sector, these may lead to $h \to \ell_1 \ell_2 \gamma$ decays,
where the presence of a hard photon in the decay would allow experiments to distinguish it from $h \to \ell \ell'$ mediated by Yukawa couplings.

As discussed before, the strength of Higgs dipole-type interactions is fixed by the strength of the corresponding dipole interaction between fermions and a gauge boson.
Therefore the constraint on the Higgs coupling will come from dipole mediated $\ell_1 \to \ell_2 \gamma$ decays.
In the limit where the leptons are massless, the width of the latter is given by:
 \beq
 \Gamma(\ell_1 \to \ell_2 \gamma) = {e^2 m_{\ell_1}^4 m_{\ell_2} \over 4 \pi v^4}
 \left(\left| [d_{Ae}]_{ \ell_1 \ell_2} \right|^2 +  \left| [d_{Ae}]_{ \ell_2 \ell_1} \right|^2 \right),
 \eeq
where we summed over the $\ell_1^+ \ell_2^-$ and $\ell_1^- \ell_2^+$ decay modes.
Using the experimental results from \tref{brlims_rad_l_fv_processes},
we get the following constraints on the dipole couplings:
\bea
\label{eq:dipole_constraints}
\sqrt{\left| [d_{Ae}]_{ e \mu} \right|^2 +  \left| [d_{Ae}]_{ \mu e} \right|^2}  & \leq & 1.2 \times 10^{-6},
\nnl
\sqrt{\left| [d_{Ae}]_{ e \tau} \right|^2 +  \left| [d_{Ae}]_{ \tau e} \right|^2}  & \leq &  2.6 \times 10^{-3},
\nnl
\sqrt{\left| [d_{Ae}]_{ \mu \tau} \right|^2 +  \left| [d_{Ae}]_{ \tau \mu} \right|^2}  & \leq &    2.1 \times 10^{-4}.
\eea
The dipole mediated $h \to \ell_1 \ell_2 \gamma$ decay width is given by:
\beq
\Gamma(h \to \ell_1 \ell_2 \gamma) = {e^2 m_h^5 m_{\ell_1} m_{\ell_2} \over 384 \pi^3 v^6}
 \left(\left| [d_{Ae}]_{ \ell_1 \ell_2} \right|^2 +  \left| [d_{Ae}]_{ \ell_2 \ell_1} \right|^2 \right),
\eeq
where we summed over the $\ell_1^+ \ell_2^-$ and $\ell_1^- \ell_2^+$ decay modes.
Given \eref{dipole_constraints}, the branching fractions for dipole mediated  $h \to \ell_1 \ell_2 \gamma$ decays are constrained\footnote{%
Of course, the process $h \to \ell_1 \ell_2 \gamma$ can occur with a larger branching fraction if it is mediated by off-diagonal Yukawa couplings and the photon is emitted by one of the final-state leptons.}  
 as:
\bea
{\rm Br}(h \to \mu e \gamma) &\leq & 1.9 \times 10^{-23},
\nnl
{\rm Br}(h \to \tau e  \gamma) &\leq & 1.7 \times 10^{-15},
\nnl
{\rm Br}(h \to \tau \mu  \gamma) &\leq & 2.3 \times 10^{-15}.
\eea
Unlike for Yukawa mediated 2-body decays,
this time the decays with $\tau$ in the final states are constrained to be extremely rare.
As long as the EFT framework is adequate for describing Higgs decays,
there is no prospect of observing the dipole mediated LFV decays at the LHC or the future 100~TeV collider~\cite{Arkani-Hamed:2015vfh}.

\subsection{$h \to \ell \ell' Z$}

Another process that can be generated by dipole-type interactions in \eref{hffdv} is $h \to \ell \ell' Z$.
An analogous calculation as in the previous section yields the decay width:
\bea &
\Gamma(h \to \ell_1 \ell_2 Z) =
 {(g_L^2 + g_Y^2)m_{\ell_1} m_{\ell_2} \over 96 \pi^3 v^6 m_h^3}
  \left(\left| [d_{Ze}]_{ \ell_1 \ell_2} \right|^2 +  \left| [d_{Ze}]_{ \ell_2 \ell_1} \right|^2 \right)
  \nnl  &\times
\int_0^{(m_h - m_Z)^2} dq^2
\sqrt{m_h^4 + (m_Z^2 - q^2)^2 -  2 m_h^2 (m_Z^2 + q^2)}
\left(m_h^4 + m_Z^4 + m_Z^2 q^2 + q^4 -     2 m_h^2 (m_Z^2 + q^2) \right),
\nnl
\eea
where we summed over the $\ell_1^+ \ell_2^-$ and $\ell_1^- \ell_2^+$ decay modes.
After evaluation of the integral we get the branching fractions:
\bea
\label{eq:zdipole_constraints}
 {\rm Br}(h \to \mu e Z)  & = & 1.4 \times 10^{-12} \left( \left| [d_{Ze}]_{ e \mu} \right|^2 +  \left| [d_{Ze}]_{ \mu e} \right|^2 \right),
\nnl
 {\rm Br}(h \to \tau e Z)   & = &  2.4 \times 10^{-11} \left( \left| [d_{Ze}]_{ e \tau} \right|^2 +  \left| [d_{Ze}]_{ \tau e} \right|^2 \right),
\nnl
 {\rm Br}(h \to \tau \mu Z)   &= &    4.9 \times 10^{-9} \left( \left| [d_{Ze}]_{ \mu \tau} \right|^2 +  \left| [d_{Ze}]_{ \tau \mu} \right|^2 \right).
\eea

	\begin{table}[tb]
		\centering
		\caption{Experimental $95\%$ CL upper limits on the branching fraction $Br$ for LFV $Z$ boson decays.}
		\label{tab:zlfv}

		\renewcommand*{\arraystretch}{1.3}

		\begin{tabular}{ c | c | c }
			\hline\hline
			Process            & Upper limits on $Br$ & Ref./Exp. \\
			\hline
			$Z^0 \to \mu e$    & \num{2.5e-6} & \cite{Abreu:1996mj} (DELPHI) \\
			                   & \num{1.7e-6} & \cite{Akers:1995gz} (OPAL) \\
			                   & \num{7.5e-7} & \cite{Aad:2014bca}  (ATLAS) \\
			\cline{2-3}
			$Z^0 \to \tau e$   & \num{2.2e-5} & \cite{Abreu:1996mj} (DELPHI) \\
			                   & \num{9.8e-6} & \cite{Akers:1995gz} (OPAL) \\
			\cline{2-3}
			$Z^0 \to \tau \mu$ & \num{1.2e-5} & \cite{Abreu:1996mj} (DELPHI) \\
			                   & \num{1.7e-5} & \cite{Akers:1995gz} (OPAL) \\
			\hline\hline
		\end{tabular}
	\end{table}

Constraints on the parameters $d_{Ze}$ come from experimental limits on LFV $Z$ boson decays summarized in \tref{zlfv}.
The dipole mediated partial decay width is given by:
\beq
\Gamma(Z \to \ell_1 \ell_2) =
 {(g_L^2 + g_Y^2) m_Z^3 m_{\ell_1} m_{\ell_2} \over 6 \pi v^4}
  \left(\left| [d_{Ze}]_{ \ell_1 \ell_2} \right|^2 +  \left| [d_{Ze}]_{ \ell_2 \ell_1} \right|^2 \right),
\eeq
where we summed over the $\ell_1^+ \ell_2^-$ and $\ell_1^- \ell_2^+$ decay modes.
This results in the following constraints on the dipole couplings:
\bea
\label{eq:zdipole_constraints_tree}
\sqrt{\left| [d_{Ze}]_{ e \mu} \right|^2 +  \left| [d_{Ze}]_{ \mu e} \right|^2}  & \leq & 76 ,
\nnl
\sqrt{\left| [d_{Ze}]_{ e \tau} \right|^2 +  \left| [d_{Ze}]_{ \tau e} \right|^2}  & \leq & 67,
\nnl
\sqrt{\left| [d_{Ze}]_{ \mu \tau} \right|^2 +  \left| [d_{Ze}]_{ \tau \mu} \right|^2}  & \leq & 5.2.
\eea
Stronger constraints on these couplings are obtained through their loop contributions to radiative lepton decays~\cite{Crivellin:2013hpa,Pruna:2014asa}.
At one loop one finds:
\beq
\Gamma(\ell_1 \to \ell_2 \gamma) =
 {m_{\ell_1}^4 m_{\ell_2} e^2 m_Z^2(g_L^2 + g_Y^2) \over 1024 \pi^5 v^6} \left(3 - 6 c_\theta^2 + 4 c_\theta^2 \log c_\theta^2 \right) ^2
  \left(\left| [d_{Ze}]_{ \ell_1 \ell_2} \right|^2 +  \left| [d_{Ze}]_{ \ell_2 \ell_1} \right|^2 \right).
\eeq
Using the experimental results from \tref{brlims_rad_l_fv_processes},
and assuming no cancellations between the tree-level $d_{Ae}$ and the one-loop contribution from $Z$ dipole,
we get the following constraints on $d_{Ze}$:
\bea
\label{eq:zdipole_constraints_loop}
\sqrt{\left| [d_{Ze}]_{ e \mu} \right|^2 +  \left| [d_{Ze}]_{ \mu e} \right|^2}  & \leq & 2.7 \times 10^{-4},
\nnl
\sqrt{\left| [d_{Ze}]_{ e \tau} \right|^2 +  \left| [d_{Ze}]_{ \tau e} \right|^2}  & \leq &  0.63,
\nnl
\sqrt{\left| [d_{Ze}]_{ \mu \tau} \right|^2 +  \left| [d_{Ze}]_{ \tau \mu} \right|^2}  & \leq &    5.1 \times 10^{-2}.
\eea
This translates to the constraints on the branching fractions:
\bea
{\rm Br}(h \to \mu e Z) &\leq & 1.1 \times 10^{-19},
\nnl
{\rm Br}(h \to \tau e  Z) &\leq & 9.6 \times 10^{-12},
\nnl
{\rm Br}(h \to \tau \mu  Z) &\leq & 1.3 \times 10^{-11}.
\eea
The suppression is slightly smaller than for the decays with a photon in the final state,
however observing decays with this low branching fraction is impossible at the LHC or at the 100~TeV collider.

The same process (though with a different helicity structure for the final state fermions) can also be generated by vertex-type couplings in \eref{hvff}.
Implementing the relevant vertices in FeynRules~\cite{Christensen:2008py,Alloul:2013bka} and calculating the decay width numerically in aMC@NLO~\cite{Alwall:2014hca} one finds:
\beq
{\rm  Br}(h \to \ell_1 \ell_2 Z) \approx 7.0 \times 10^{-5}
  \left(\left| [\delta g^{Ze}_L]_{ \ell_1 \ell_2} \right|^2
  + \left| [\delta g^{Ze}_L]_{ \ell_2 \ell_1} \right|^2
+    \left| [\delta g^{Ze}_R]_{\ell_1 \ell_2} \right|^2
+   \left| [\delta g^{Ze}_R]_{\ell_2  \ell_1} \right|^2\right).
\eeq
Again, the off-diagonal vertex corrections are constrained by LFV $Z$ boson decays.
The decay width is:
\beq
\Gamma(Z  \to \ell_1 \ell_2 ) = {(g_L^2 + g_Y^2) m_Z \over 24 \pi}
  \left(\left| [\delta g^{Ze}_L]_{ \ell_1 \ell_2} \right|^2
  + \left| [\delta g^{Ze}_L]_{ \ell_2 \ell_1} \right|^2
+    \left| [\delta g^{Ze}_R]_{\ell_1 \ell_2} \right|^2
+   \left| [\delta g^{Ze}_R]_{\ell_2  \ell_1} \right|^2\right).
\eeq
Then the experimental constraints in \tref{zlfv} imply:
\bea
\sqrt{ \left| [\delta g^{Ze}_L]_{\mu e} \right|^2
     + \left| [\delta g^{Ze}_L]_{e \mu} \right|^2
     + \left| [\delta g^{Ze}_R]_{\mu e} \right|^2
     + \left| [\delta g^{Ze}_R]_{e \mu} \right|^2 } & \leq & 1.7 \times 10^{-3},
\nnl
\sqrt{ \left| [\delta g^{Ze}_L]_{\tau e} \right|^2
     + \left| [\delta g^{Ze}_L]_{e \tau} \right|^2
     + \left| [\delta g^{Ze}_R]_{\tau e} \right|^2
     + \left| [\delta g^{Ze}_R]_{e \tau} \right|^2 } & \leq & 6.1 \times 10^{-3},
\nnl
\sqrt{ \left| [\delta g^{Ze}_L]_{\tau \mu} \right|^2
     + \left| [\delta g^{Ze}_L]_{\mu \tau} \right|^2
     + \left| [\delta g^{Ze}_R]_{\tau \mu} \right|^2
     + \left| [\delta g^{Ze}_R]_{\mu \tau} \right|^2 } & \leq & 6.8 \times 10^{-3}.
\eea

	\begin{table}[tb]
		\centering
		\caption{Experimental $90\%$ CL. upper limits on the branching fraction $Br$ for 4-lepton flavour-violating processes.}
		\label{tab:brlims_4l_fv_processes}

		\renewcommand*{\arraystretch}{1.3}

		\begin{tabular}{ c | c | c }
			\hline\hline
			Process             & Upper limits on $Br$ & Ref./Exp. \\
			\hline
			$\mu \to 3e$        & \num{1.0e-12} & \cite{Bellgardt:1987du} (SINDRUM) \\
			$\tau \to 3e$       & \num{2.7e-8}  & \cite{Hayasaka:2010np} (Belle) \\
			$\tau \to 3\mu$     & \num{2.1e-8}  & \cite{Hayasaka:2010np} (Belle) \\
			\hline\hline
		\end{tabular}
	\end{table}

Again stronger constraints arise through 1-loop contributions to LFV lepton decays,
for which the experimental limits are collected in \tref{brlims_4l_fv_processes}.
Assuming no cancellations with tree-level contributions of 4-fermion operators,
one obtains the bounds~\cite{Pruna:2015jhf}: 
\bea
\sqrt{ \left| [\delta g^{Ze}_L]_{\mu e} \right|^2
     + \left| [\delta g^{Ze}_L]_{e \mu} \right|^2
     + \left| [\delta g^{Ze}_R]_{\mu e} \right|^2
     + \left| [\delta g^{Ze}_R]_{e \mu} \right|^2 } & \leq & 1.5 \times 10^{-6},
\nnl
\sqrt{ \left| [\delta g^{Ze}_L]_{\tau e} \right|^2
     + \left| [\delta g^{Ze}_L]_{e \tau} \right|^2
     + \left| [\delta g^{Ze}_R]_{\tau e} \right|^2
     + \left| [\delta g^{Ze}_R]_{e \tau} \right|^2 } & \leq & 8.4 \times 10^{-4},
\nnl
\sqrt{ \left| [\delta g^{Ze}_L]_{\tau \mu} \right|^2
     + \left| [\delta g^{Ze}_L]_{\mu \tau} \right|^2
     + \left| [\delta g^{Ze}_R]_{\tau \mu} \right|^2
     + \left| [\delta g^{Ze}_R]_{\mu \tau} \right|^2 } & \leq & 5.9 \times 10^{-4}.
\eea
This translates to the following bounds on the branching fractions:
\bea
{\rm Br}(h \to \mu e Z) &\leq & 1.6 \times 10^{-16},
\nnl
{\rm Br}(h \to \tau e Z) &\leq & 4.9 \times 10^{-11},
\nnl
{\rm Br}(h \to \tau \mu Z) &\leq & 2.4 \times 10^{-11}.
\eea
The bounds are somewhat weaker than for the dipole mediated Higgs decays,
however the suppression is still too much for any realistic prospects of experimental detection.

\subsection{$t \to h q V$}

Finally, we consider flavor-violating 3-body decays of the top quark mediated by dipole-type operators:
$t \to h q V$, where $V$ is a photon or a gluon.
We have implemented the $\bar{t} \sigma^{\mu\nu} c V_{\mu\nu}$ and $h \bar{t} \sigma^{\mu\nu} c V_{\mu\nu}$ vertices in FeynRules, and calculated the decay width numerically in aMC@NLO. 
We find:
\bea
{ {\rm Br}(t \to q V h) \over {\rm Br}(t \to q V) } \approx 4.4 \times 10^{-8}.
\eea
The current best constraints on ${\rm Br}(t \to q \gamma)$ come from searches for anomalous top production at the LHC.
For the dipole couplings to photons the strongest limits come from the CMS experiment~\cite{Khachatryan:2015att}.
They translate to the following limits on the branching fractions:
\bea
\label{eq:tqga_limits}
{\rm Br}(t \to u \gamma) &\leq & 1.3 \times 10^{-4} ,
\nnl
{\rm Br}(t \to c \gamma) &\leq & 1.7 \times 10^{-3} .
\eea
For $t \to u \gamma$, even stronger limits can be placed due to the dipole contributions to the neutron electric dipole moment~\cite{Khatibi:2015aal}, though these constraints do no apply when the dipole couplings are parity conserving.
For the dipole couplings to the gluon, the strongest limits come from the ATLAS experiments~\cite{Aad:2015gea}:
\bea
\label{eq:tqg_limits}
{\rm Br}(t \to u g) &\leq & 4.0 \times 10^{-5} ,
\nnl
{\rm Br}(t \to c g) &\leq & 1.7 \times 10^{-4} .
\eea
Limits on the flavor violating dipole top couplings from $t \gamma$ production at the LHC~\cite{Durieux:2014xla} are currently weaker. 
The experimental bounds in \eref{tqga_limits} and \eref{tqg_limits} translate to the following constraints on dipole mediated top decays with the Higgs:
\bea
\label{eq:br_tqg}
{\rm Br}(t \to u \gamma h) & \leq & 5.7 \times 10^{-12},
\nnl
{\rm Br}(t \to c \gamma h) & \leq & 7.5 \times 10^{-11},
\nnl
{\rm Br}(t \to u g h) & \leq & 1.8 \times 10^{-12},
\nnl
{\rm Br}(t \to c g h) & \leq & 7.5 \times 10^{-12} .
\eea
As in the case of $h \to \ell^+_1 \ell^-_2  \gamma$ decays, the branching fraction for $t \to q \gamma/g h$ may be larger than the limits in \eref{br_tqg} if the process is mediated by off-diagonal Yukawa couplings and the photon or gluon is emitted from the final state quark.  
Although  the top production cross section is larger than that of the Higgs boson ($\cO(1)$~nb at the 14~TeV LHC,
and a factor of 30 larger at 100~TeV, which corresponds to  $\cO(10^{11})$ $t\bar t$ pairs  in the future 100~TeV collider~\cite{Arkani-Hamed:2015vfh}), the limits in \eref{br_tqg} leave little room for observing the dipole mediated top quark decays. 
A more realistic probe of flavor violating dipole-type Higgs interactions may be offered by the production processes  $p p \to h t j$ and  $p p \to h t \gamma$.\footnote{We thank the authors of Ref.~\cite{Durieux:2014xla}  for pointing this out to us.}
 



\section{Conclusions}

In this paper we discussed the possibility of observing at the LHC exotic Higgs decays that violate lepton or quark flavor.
Our study was done in the context of an EFT which describes the effective interactions of the Higgs boson with other SM particles after heavy particles from beyond the SM have been integrated out.
In this context, the possibility of a significant rate of 2-body decays such as
$h \to \mu \tau$, $h \to e \tau$, and $t \to c h$ was pointed out in the previous literature.
Our analysis confirms and updates these conclusions.

We also studied the possibility of exotic 3-body decays involving the Higgs boson.
Here, our conclusions are largely negative.
The existing precision constraints imply that the rate of such 3-body processes must be prohibitively small and cannot be observed in colliders in the foreseeable future.
This is an important and robust conclusion that can derived in the EFT framework.
Conversely, if such 3-body processes are observed, this would signal a breakdown of the EFT description we used regarding Higgs decay processes.
Such a breakdown would be a harbinger of new light degrees of freedom,
or a non-linear realization of electroweak symmetry.



\section*{Acknowledgements}

AF~is supported by the ERC Advanced Grant Higgs@LHC.

\appendix

\bibliographystyle{JHEP}
\bibliography{exoeft}

\providecommand{\href}[2]{#2}\begingroup\raggedright\begin{thebibliography}{10}

\bibitem{Buchmuller:1985jz}
W.~Buchmuller and D.~Wyler, {\it {Effective Lagrangian Analysis of New
  Interactions and Flavor Conservation}},  {\em Nucl. Phys.} {\bf B268} (1986)
  621--653.

\bibitem{Grzadkowski:2010es}
B.~Grzadkowski, M.~Iskrzynski, M.~Misiak, and J.~Rosiek, {\it {Dimension-Six
  Terms in the Standard Model Lagrangian}},  {\em JHEP} {\bf 10} (2010) 085,
  [\href{http://arxiv.org/abs/1008.4884}{{\tt arXiv:1008.4884}}].

\bibitem{Alonso:2013hga}
R.~Alonso, E.~E. Jenkins, A.~V. Manohar, and M.~Trott, {\it {Renormalization
  Group Evolution of the Standard Model Dimension Six Operators III: Gauge
  Coupling Dependence and Phenomenology}},  {\em JHEP} {\bf 1404} (2014) 159,
  [\href{http://arxiv.org/abs/1312.2014}{{\tt arXiv:1312.2014}}].

\bibitem{Strassler:2006ri}
M.~J. Strassler and K.~M. Zurek, {\it {Discovering the Higgs through
  highly-displaced vertices}},  {\em Phys. Lett.} {\bf B661} (2008) 263--267,
  [\href{http://arxiv.org/abs/hep-ph/0605193}{{\tt arXiv:hep-ph/0605193}}].

\bibitem{Strassler:2006im}
M.~J. Strassler and K.~M. Zurek, {\it {Echoes of a hidden valley at hadron
  colliders}},  {\em Phys. Lett.} {\bf B651} (2007) 374--379,
  [\href{http://arxiv.org/abs/hep-ph/0604261}{{\tt arXiv:hep-ph/0604261}}].

\bibitem{Gopalakrishna:2008dv}
S.~Gopalakrishna, S.~Jung, and J.~D. Wells, {\it {Higgs boson decays to four
  fermions through an abelian hidden sector}},  {\em Phys. Rev.} {\bf D78}
  (2008) 055002, [\href{http://arxiv.org/abs/0801.3456}{{\tt
  arXiv:0801.3456}}].

\bibitem{Davoudiasl:2012ag}
H.~Davoudiasl, H.-S. Lee, and W.~J. Marciano, {\it {'Dark' Z implications for
  Parity Violation, Rare Meson Decays, and Higgs Physics}},  {\em Phys. Rev.}
  {\bf D85} (2012) 115019, [\href{http://arxiv.org/abs/1203.2947}{{\tt
  arXiv:1203.2947}}].

\bibitem{Davoudiasl:2013aya}
H.~Davoudiasl, H.-S. Lee, I.~Lewis, and W.~J. Marciano, {\it {Higgs Decays as a
  Window into the Dark Sector}},  {\em Phys. Rev.} {\bf D88} (2013), no.~1
  015022, [\href{http://arxiv.org/abs/1304.4935}{{\tt arXiv:1304.4935}}].

\bibitem{Huang:2013ima}
J.~Huang, T.~Liu, L.-T. Wang, and F.~Yu, {\it {Supersymmetric Exotic Decays of
  the 125 GeV Higgs Boson}},  {\em Phys. Rev. Lett.} {\bf 112} (2014), no.~22
  221803, [\href{http://arxiv.org/abs/1309.6633}{{\tt arXiv:1309.6633}}].

\bibitem{Gonzalez-Alonso:2014rla}
M.~Gonzalez-Alonso and G.~Isidori, {\it {The $h \to 4\ell$ spectrum at low
  $m_{34}$: Standard Model vs. light New Physics}},  {\em Phys. Lett.} {\bf
  B733} (2014) 359--365, [\href{http://arxiv.org/abs/1403.2648}{{\tt
  arXiv:1403.2648}}].

\bibitem{Falkowski:2014ffa}
A.~Falkowski and R.~Vega-Morales, {\it {Exotic Higgs decays in the golden
  channel}},  {\em JHEP} {\bf 12} (2014) 037,
  [\href{http://arxiv.org/abs/1405.1095}{{\tt arXiv:1405.1095}}].

\bibitem{Curtin:2013fra}
D.~Curtin, R.~Essig, S.~Gori, P.~Jaiswal, A.~Katz, et~al., {\it {Exotic Decays
  of the 125 GeV Higgs Boson}},  {\em Phys. Rev.} {\bf D90} (2014), no.~7
  075004, [\href{http://arxiv.org/abs/1312.4992}{{\tt arXiv:1312.4992}}].

\bibitem{Blankenburg:2012ex}
G.~Blankenburg, J.~Ellis, and G.~Isidori, {\it {Flavour-Changing Decays of a
  125 GeV Higgs-like Particle}},  {\em Phys. Lett.} {\bf B712} (2012) 386--390,
  [\href{http://arxiv.org/abs/1202.5704}{{\tt arXiv:1202.5704}}].

\bibitem{Harnik:2012pb}
R.~Harnik, J.~Kopp, and J.~Zupan, {\it {Flavor Violating Higgs Decays}},  {\em
  JHEP} {\bf 03} (2013) 026, [\href{http://arxiv.org/abs/1209.1397}{{\tt
  arXiv:1209.1397}}].

\bibitem{DiazCruz:1999xe}
J.~L. Diaz-Cruz and J.~J. Toscano, {\it {Lepton flavor violating decays of
  Higgs bosons beyond the standard model}},  {\em Phys. Rev.} {\bf D62} (2000)
  116005, [\href{http://arxiv.org/abs/hep-ph/9910233}{{\tt
  arXiv:hep-ph/9910233}}].

\bibitem{Chatrchyan:2014tja}
{\bf CMS} Collaboration, S.~Chatrchyan et~al., {\it {Search for invisible
  decays of Higgs bosons in the vector boson fusion and associated ZH
  production modes}},  {\em Eur.Phys.J.} {\bf C74} (2014), no.~8 2980,
  [\href{http://arxiv.org/abs/1404.1344}{{\tt arXiv:1404.1344}}].

\bibitem{Gupta:2014rxa}
R.~S. Gupta, A.~Pomarol, and F.~Riva, {\it {BSM Primary Effects}},  {\em Phys.
  Rev.} {\bf D91} (2015), no.~3 035001,
  [\href{http://arxiv.org/abs/1405.0181}{{\tt arXiv:1405.0181}}].

\bibitem{HXSWGbasis}
{\bf LHC Higgs Cross Section Working Group 2} Collaboration, {\it {Higgs Basis:
  Proposal for an EFT basis choice for LHC HXSWG}},  tech. rep., 2015.

\bibitem{deLima:2015pqa}
L.~de~Lima, C.~S. Machado, R.~D. Matheus, and L.~A.~F. do~Prado, {\it {Higgs
  Flavor Violation as a Signal to Discriminate Models}},  {\em JHEP} {\bf 11}
  (2015) 074, [\href{http://arxiv.org/abs/1501.06923}{{\tt arXiv:1501.06923}}].

\bibitem{Khachatryan:2015kon}
{\bf CMS} Collaboration, V.~Khachatryan et~al., {\it {Search for
  Lepton-Flavour-Violating Decays of the Higgs Boson}},  {\em Phys. Lett.} {\bf
  B749} (2015) 337--362, [\href{http://arxiv.org/abs/1502.07400}{{\tt
  arXiv:1502.07400}}].

\bibitem{Aad:2015gha}
{\bf ATLAS} Collaboration, G.~Aad et~al., {\it {Search for
  lepton-flavour-violating $H\to\mu\tau$ decays of the Higgs boson with the
  ATLAS detector}},  \href{http://arxiv.org/abs/1508.03372}{{\tt
  arXiv:1508.03372}}.

\bibitem{Bhattacherjee:2015sia}
B.~Bhattacherjee, S.~Chakraborty, and S.~Mukherjee, {\it {$H \rightarrow \tau
  \mu$ and excess in $t\bar{t}H$: Connecting the dots in the hope for the first
  glimpse of BSM Higgs signal}},  \href{http://arxiv.org/abs/1505.02688}{{\tt
  arXiv:1505.02688}}.

\bibitem{Crivellin:2014cta}
A.~Crivellin, M.~Hoferichter, and M.~Procura, {\it {Improved predictions for
  $\mu\to e$ conversion in nuclei and Higgs-induced lepton flavor violation}},
  {\em Phys. Rev.} {\bf D89} (2014) 093024,
  [\href{http://arxiv.org/abs/1404.7134}{{\tt arXiv:1404.7134}}].

\bibitem{Davidson:2016utf}
S.~Davidson, {\it {Mu to e gamma in the 2 Higgs Doublet Model: an exercise in
  EFT}},  \href{http://arxiv.org/abs/1601.01949}{{\tt arXiv:1601.01949}}.

\bibitem{Chang:1993kw}
D.~Chang, W.~S. Hou, and W.-Y. Keung, {\it {Two loop contributions of flavor
  changing neutral Higgs bosons to $\mu \to e \gamma$}},  {\em Phys. Rev.} {\bf
  D48} (1993) 217--224, [\href{http://arxiv.org/abs/hep-ph/9302267}{{\tt
  arXiv:hep-ph/9302267}}].

\bibitem{Leigh:1990kf}
R.~G. Leigh, S.~Paban, and R.~M. Xu, {\it {Electric dipole moment of
  electron}},  {\em Nucl. Phys.} {\bf B352} (1991) 45--58.

\bibitem{Adam:2013mnn}
{\bf MEG} Collaboration, J.~Adam et~al., {\it {New constraint on the existence
  of the $\mu^+ \to e^+\gamma$ decay}},  {\em Phys. Rev. Lett.} {\bf 110}
  (2013) 201801, [\href{http://arxiv.org/abs/1303.0754}{{\tt
  arXiv:1303.0754}}].

\bibitem{Aubert:2009ag}
{\bf BaBar} Collaboration, B.~Aubert et~al., {\it {Searches for Lepton Flavor
  Violation in the Decays $\tau^\pm \rightarrow e^\pm \gamma$ and $\tau^\pm
  \rightarrow \mu^\pm \gamma$}},  {\em Phys. Rev. Lett.} {\bf 104} (2010)
  021802, [\href{http://arxiv.org/abs/0908.2381}{{\tt arXiv:0908.2381}}].

\bibitem{Pruna:2015jhf}
G.~M. Pruna and A.~Signer, {\it {Lepton-flavour violating decays in theories
  with dimension 6 operators}},  in {\em {Proceedings, GPU Computing in
  High-Energy Physics (GPUHEP2014)}}, 2015.
\newblock \href{http://arxiv.org/abs/1511.04421}{{\tt arXiv:1511.04421}}.

\bibitem{Falkowski:2013jya}
A.~Falkowski, D.~M. Straub, and A.~Vicente, {\it {Vector-like leptons: Higgs
  decays and collider phenomenology}},  {\em JHEP} {\bf 05} (2014) 092,
  [\href{http://arxiv.org/abs/1312.5329}{{\tt arXiv:1312.5329}}].

\bibitem{Dery:2014kxa}
A.~Dery, A.~Efrati, Y.~Nir, Y.~Soreq, and V.~Susič, {\it {Model building for
  flavor changing Higgs couplings}},  {\em Phys. Rev.} {\bf D90} (2014) 115022,
  [\href{http://arxiv.org/abs/1408.1371}{{\tt arXiv:1408.1371}}].

\bibitem{Sierra:2014nqa}
D.~Aristizabal~Sierra and A.~Vicente, {\it {Explaining the CMS Higgs flavor
  violating decay excess}},  {\em Phys. Rev.} {\bf D90} (2014), no.~11 115004,
  [\href{http://arxiv.org/abs/1409.7690}{{\tt arXiv:1409.7690}}].

\bibitem{Crivellin:2015mga}
A.~Crivellin, G.~D'Ambrosio, and J.~Heeck, {\it {Explaining
  $h\to\mu^\pm\tau^\mp$, $B\to K^* \mu^+\mu^-$ and $B\to K \mu^+\mu^-/B\to K
  e^+e^-$ in a two-Higgs-doublet model with gauged $L_\mu-L_\tau$}},  {\em
  Phys. Rev. Lett.} {\bf 114} (2015) 151801,
  [\href{http://arxiv.org/abs/1501.00993}{{\tt arXiv:1501.00993}}].

\bibitem{Crivellin:2015lwa}
A.~Crivellin, G.~D'Ambrosio, and J.~Heeck, {\it {Addressing the LHC flavor
  anomalies with horizontal gauge symmetries}},  {\em Phys. Rev.} {\bf D91}
  (2015), no.~7 075006, [\href{http://arxiv.org/abs/1503.03477}{{\tt
  arXiv:1503.03477}}].

\bibitem{Crivellin:2015hha}
A.~Crivellin, J.~Heeck, and P.~Stoffer, {\it {A perturbed lepton-specific
  two-Higgs-doublet model facing experimental hints for physics beyond the
  Standard Model}},  {\em Phys. Rev. Lett.} {\bf 116} (2016), no.~8 081801,
  [\href{http://arxiv.org/abs/1507.07567}{{\tt arXiv:1507.07567}}].

\bibitem{Varzielas:2015joa}
I.~de~Medeiros~Varzielas, O.~Fischer, and V.~Maurer, {\it {$ {\mathbb{A}}_4 $
  symmetry at colliders and in the universe}},  {\em JHEP} {\bf 08} (2015) 080,
  [\href{http://arxiv.org/abs/1504.03955}{{\tt arXiv:1504.03955}}].

\bibitem{Aloni:2015wvn}
D.~Aloni, Y.~Nir, and E.~Stamou, {\it {Large $BR(h \to \tau\mu)$ in the MSSM}},
   \href{http://arxiv.org/abs/1511.00979}{{\tt arXiv:1511.00979}}.

\bibitem{He:2015rqa}
X.-G. He, J.~Tandean, and Y.-J. Zheng, {\it {Higgs decay $h \to \mu\tau$ with
  minimal flavor violation}},  {\em JHEP} {\bf 09} (2015) 093,
  [\href{http://arxiv.org/abs/1507.02673}{{\tt arXiv:1507.02673}}].

\bibitem{Baek:2015mea}
S.~Baek and K.~Nishiwaki, {\it {Leptoquark explanation of $h \to \mu\tau$ and
  muon $(g-2)$}},  \href{http://arxiv.org/abs/1509.07410}{{\tt
  arXiv:1509.07410}}.

\bibitem{Dorsner:2015mja}
I.~Doršner, S.~Fajfer, A.~Greljo, J.~F. Kamenik, N.~Košnik, and
  I.~Nišandžic, {\it {New Physics Models Facing Lepton Flavor Violating Higgs
  Decays at the Percent Level}},  {\em JHEP} {\bf 06} (2015) 108,
  [\href{http://arxiv.org/abs/1502.07784}{{\tt arXiv:1502.07784}}].

\bibitem{Bizot:2015qqo}
N.~Bizot, S.~Davidson, M.~Frigerio, and J.-L. Kneur, {\it {Two Higgs doublets
  to explain the excesses $pp\rightarrow \gamma\gamma(750\ {\rm GeV})$ and $h
  \to \tau^\pm \mu^\mp$}},  \href{http://arxiv.org/abs/1512.08508}{{\tt
  arXiv:1512.08508}}.

\bibitem{Buschmann:2016uzg}
M.~Buschmann, J.~Kopp, J.~Liu, and X.-P. Wang, {\it {New Signatures of Flavor
  Violating Higgs Couplings}},  \href{http://arxiv.org/abs/1601.02616}{{\tt
  arXiv:1601.02616}}.

\bibitem{Greljo:2014dka}
A.~Greljo, J.~F. Kamenik, and J.~Kopp, {\it {Disentangling Flavor Violation in
  the Top-Higgs Sector at the LHC}},  {\em JHEP} {\bf 07} (2014) 046,
  [\href{http://arxiv.org/abs/1404.1278}{{\tt arXiv:1404.1278}}].

\bibitem{Aad:2015pja}
{\bf ATLAS} Collaboration, G.~Aad et~al., {\it {Search for flavour-changing
  neutral current top quark decays $t\to Hq$ in $pp$ collisions at $\sqrt{s}=8$
  TeV with the ATLAS detector}},  {\em JHEP} {\bf 12} (2015) 061,
  [\href{http://arxiv.org/abs/1509.06047}{{\tt arXiv:1509.06047}}].

\bibitem{Aad:2014dya}
{\bf ATLAS} Collaboration, G.~Aad et~al., {\it {Search for top quark decays $t
  \to qH$ with $H \to \gamma\gamma$ using the ATLAS detector}},  {\em JHEP}
  {\bf 06} (2014) 008, [\href{http://arxiv.org/abs/1403.6293}{{\tt
  arXiv:1403.6293}}].

\bibitem{CMS:2014qxa}
{\bf CMS} Collaboration, {\it {Combined multilepton and diphoton limit on t to
  cH}},  {\em CMS-PAS-HIG-13-034} (2014).

\bibitem{Gorbahn:2014sha}
M.~Gorbahn and U.~Haisch, {\it {Searching for $t \to c(u)h$ with dipole
  moments}},  {\em JHEP} {\bf 06} (2014) 033,
  [\href{http://arxiv.org/abs/1404.4873}{{\tt arXiv:1404.4873}}].

\bibitem{Arkani-Hamed:2015vfh}
N.~Arkani-Hamed, T.~Han, M.~Mangano, and L.-T. Wang, {\it {Physics
  Opportunities of a 100 TeV Proton-Proton Collider}},
  \href{http://arxiv.org/abs/1511.06495}{{\tt arXiv:1511.06495}}.

\bibitem{Abreu:1996mj}
{\bf DELPHI} Collaboration, P.~Abreu et~al., {\it {Search for lepton flavor
  number violating $Z^0$ decays}},  {\em Z. Phys.} {\bf C73} (1997) 243--251.

\bibitem{Akers:1995gz}
{\bf OPAL} Collaboration, R.~Akers et~al., {\it {A Search for lepton flavor
  violating $Z^0$ decays}},  {\em Z. Phys.} {\bf C67} (1995) 555--564.

\bibitem{Aad:2014bca}
{\bf ATLAS} Collaboration, G.~Aad et~al., {\it {Search for the lepton flavor
  violating decay $Z \rightarrow e \mu$ in $pp$ collisions at $\sqrt{s}=8$ TeV
  with the ATLAS detector}},  {\em Phys. Rev.} {\bf D90} (2014), no.~7 072010,
  [\href{http://arxiv.org/abs/1408.5774}{{\tt arXiv:1408.5774}}].

\bibitem{Crivellin:2013hpa}
A.~Crivellin, S.~Najjari, and J.~Rosiek, {\it {Lepton Flavor Violation in the
  Standard Model with general Dimension-Six Operators}},  {\em JHEP} {\bf 04}
  (2014) 167, [\href{http://arxiv.org/abs/1312.0634}{{\tt arXiv:1312.0634}}].

\bibitem{Pruna:2014asa}
G.~M. Pruna and A.~Signer, {\it {The $\mu\to e\gamma$ decay in a systematic
  effective field theory approach with dimension 6 operators}},  {\em JHEP}
  {\bf 10} (2014) 14, [\href{http://arxiv.org/abs/1408.3565}{{\tt
  arXiv:1408.3565}}].

\bibitem{Christensen:2008py}
N.~D. Christensen and C.~Duhr, {\it {FeynRules - Feynman rules made easy}},
  {\em Comput.Phys.Commun.} {\bf 180} (2009) 1614--1641,
  [\href{http://arxiv.org/abs/0806.4194}{{\tt arXiv:0806.4194}}].

\bibitem{Alloul:2013bka}
A.~Alloul, N.~D. Christensen, C.~Degrande, C.~Duhr, and B.~Fuks, {\it
  {FeynRules 2.0 - A complete toolbox for tree-level phenomenology}},  {\em
  Comput. Phys. Commun.} {\bf 185} (2014) 2250--2300,
  [\href{http://arxiv.org/abs/1310.1921}{{\tt arXiv:1310.1921}}].

\bibitem{Alwall:2014hca}
J.~Alwall, R.~Frederix, S.~Frixione, V.~Hirschi, F.~Maltoni, O.~Mattelaer,
  H.~S. Shao, T.~Stelzer, P.~Torrielli, and M.~Zaro, {\it {The automated
  computation of tree-level and next-to-leading order differential cross
  sections, and their matching to parton shower simulations}},  {\em JHEP} {\bf
  07} (2014) 079, [\href{http://arxiv.org/abs/1405.0301}{{\tt
  arXiv:1405.0301}}].

\bibitem{Bellgardt:1987du}
{\bf SINDRUM} Collaboration, U.~Bellgardt et~al., {\it {Search for the decay
  $\mu^+ \rightarrow e^+ e^+ e^-$}},  {\em Nucl. Phys.} {\bf B299} (1988) 1.

\bibitem{Hayasaka:2010np}
{\bf Belle} Collaboration, K.~Hayasaka, K.~Inami, Y.~Miyazaki, K.~Arinstein,
  V.~Aulchenko, et~al., {\it {Search for Lepton Flavor Violating Tau Decays
  into Three Leptons with 719 Million Produced Tau+Tau- Pairs}},  {\em Phys.
  Lett.} {\bf B687} (2010) 139--143,
  [\href{http://arxiv.org/abs/1001.3221}{{\tt arXiv:1001.3221}}].

\bibitem{Khachatryan:2015att}
{\bf CMS} Collaboration, V.~Khachatryan et~al., {\it {Search for anomalous
  single top quark production in association with a photon in pp collisions at
  $\sqrt{s} = 8$~TeV}},  \href{http://arxiv.org/abs/1511.03951}{{\tt
  arXiv:1511.03951}}.

\bibitem{Khatibi:2015aal}
S.~Khatibi and M.~M. Najafabadi, {\it {Top quark flavor changing via photon}},
  \href{http://arxiv.org/abs/1511.00220}{{\tt arXiv:1511.00220}}.

\bibitem{Aad:2015gea}
{\bf ATLAS} Collaboration, G.~Aad et~al., {\it {Search for single top-quark
  production via flavour changing neutral currents at 8 TeV with the ATLAS
  detector}},  \href{http://arxiv.org/abs/1509.00294}{{\tt arXiv:1509.00294}}.

\bibitem{Durieux:2014xla}
G.~Durieux, F.~Maltoni, and C.~Zhang, {\it {Global approach to top-quark
  flavor-changing interactions}},  {\em Phys. Rev.} {\bf D91} (2015), no.~7
  074017, [\href{http://arxiv.org/abs/1412.7166}{{\tt arXiv:1412.7166}}].

\end{thebibliography}\endgroup

\end{document}